\documentclass{article}
\usepackage{authblk}
\usepackage[utf8]{inputenc}
\usepackage{amsmath,amssymb,amsfonts,amsthm}
\usepackage{physics}
\usepackage{dsfont}
\usepackage{graphicx}
\usepackage{xcolor}
\usepackage{cancel}
\usepackage[numbers,sort&compress,square]{natbib}
\usepackage{color}
\usepackage{bbm}
\usepackage[linktocpage]{hyperref}
\usepackage{appendix}
\hypersetup{colorlinks=true,citecolor=blue,linkcolor=blue, urlcolor=blue, breaklinks=true}
\usepackage[eulergreek]{sansmath}
\usepackage{mathtools}
\usepackage{scalerel}

\usepackage{ulem}

\usepackage{bm} 
\usepackage{subfigure} 
\usepackage{environ}
\usepackage{url}
\usepackage{hyperref}
\usepackage[margin=1in]{geometry}

\usepackage{environ}
\NewEnviron{eqs}{%
\begin{equation}\begin{split}
    \BODY
\end{split}\end{equation}}

\newcommand{\Z}{{\mathbb Z}}
\newcommand{\R}{{\mathbb R}}
\newcommand{\Q}{{\mathbb Q}}

\providecommand{\indF}{\operatorname{Ind}_F}

\graphicspath{{./Figures/}}

\begin{document}

\title{Fermionic Anomalies of Finite Symmetries on Lattices}

\author[1, *]{Ameya Chavda}
\author[2, $\dagger$]{Ryohei Kobayashi}

\affil[1]{Center for Theoretical Physics, Department of Physics, Columbia University, New York, USA}
\affil[2]{Department of Applied Physics, The University of Tokyo, Tokyo 113-8656, Japan}

\renewcommand*{\thefootnote}{*}
\footnotetext[1]{Contact author: ameya.chavda@columbia.edu}
\renewcommand*{\thefootnote}{$\dagger$}
\footnotetext[1]{Contact author: ryohei.k@ap.t.u-tokyo.ac.jp}

\renewcommand{\thefootnote}{\arabic{footnote}}

\date{\today}

\maketitle

\begin{abstract}
We develop a lattice characterization of fermionic 't Hooft anomalies of finite internal symmetries in one and two spatial dimensions, formulated in terms of obstructions to symmetric short-range-entangled (SRE) states.  We consider lattice systems formed by tensor product of onsite fermionic and bosonic local degrees of freedom, and finite internal symmetry given by a central extension $\mathbb Z_2^F\to G_f\to G_b$. We extract a hierarchy of fermionic anomaly indices for a given symmetry operator in both (1+1)D and (2+1)D. In (1+1)D, an exact lattice symmetry is characterized by a pair of cohomological data $(n_2,\nu_3)$. For $G_f=G_b\times\mathbb Z_2^F$, we further show that a symmetry with trivial lattice anomaly indices $(n_2,\nu_3)$ is onsiteable and hence admits a symmetric SRE state, establishing that these indices faithfully detect the lattice anomaly. Comparing with continuum fermionic QFT, we find that exact lattice symmetries do not realize the additional $H^1(BG_b,\mathbb Z_2)$ anomaly layer in continuum QFT. In particular, for $G_b=\mathbb Z_2$, exact lattice symmetries realize only the even $\mathbb Z_4$ subgroup of the continuum $\mathbb Z_8$ classification.
In (2+1)D, we identify three successive anomaly layers of cohomological data $(n_2,n_3,\nu_4)$. We show that a nontrivial value of any layer obstructs a symmetric SRE state. For $G_f=G_b\times\mathbb Z_2^F$, it also forbids a symmetric invertible state. We also construct a non-onsite $\mathbb Z_4^F$ lattice symmetry in (2+1)D with nontrivial lattice anomaly index that forbids any symmetric invertible states with $c_-\in\Z$, therefore forbids any invertible states compatible with onsite $\Z_4^F$ symmetry. Our results motivate a systematic study of which continuum anomalies admit exact microscopic lattice realizations.

\end{abstract}

\tableofcontents

\section{Introduction}

't Hooft anomalies place robust constraints on the possible infrared behavior of quantum many-body systems. A symmetry with a nontrivial 't Hooft anomaly cannot be realized in a trivially gapped phase while preserving the symmetry. In particular, the anomaly obstructs the existence of a symmetric short-range-entangled (SRE) state, forcing the system to host a nontrivial pattern of entanglement. 
This perspective makes the anomaly relevant to microscopic lattice systems, where one can ask whether a given symmetry action admits a symmetric trivial gapped state. 
See e.g.,~\cite{else2014, seiberg2022lsm,
Seifnashri2024lsm, kawagoe2025anomaly, kapustin2025highersymmetries, seifnashri2025disentangling,
shirley2025QCA, Feng2026higherformanomalies, Feng2026onsite, Tu2026anomalies, Kapustin2025LSM, Kobayashi2026generalizedstatistics, czajka2025anomalieslatticehomotopyquantum} for lattice characterization of 't Hooft anomalies and their physical consequences.

For fermionic systems, the structure of such obstructions is richer than in bosonic systems. Fermion parity is always present as a distinguished unbreakable symmetry, and a finite internal symmetry generally takes the form of a central extension $\Z_2^F\to G_f\to G_b$.
Consequently, the operators associated with symmetry defects and their junctions can themselves be fermion-parity odd or carry Majorana zero modes, so that the anomalies contain information beyond ordinary phase ambiguity familiar in bosonic theories~\cite{FidkowskiKitaev1d, Kapustin2015spinbordism, Witten:2016cio, Tata2022anomalies,  Tong_2020, Hason_2020}. 
These features give rise to the layered structure of cohomological data to characterize the fermionic anomaly~\cite{gu2014, WangGu2020,Kapustin:2017jrc,
barkeshli2021invertible,bhardwajGaiottoKapustin2017,thorngren2019anomaliesbosonization, Delmastro2021global, brumfiel2018pontrjagindual3dimensionalspin, brumfiel2018pontrjagindual4dimensionalspin,
Bulmash2022cascade, Gaiotto2019spt}, enriching classification of bosonic anomalies by group cohomology~\cite{chen2013SPT}. 
Fermionic anomalies are ubiquitous in both quantum field theories (QFTs) and lattice systems, including free-fermion systems, and anomalous theories have been extensively studied in both continuum QFT and lattice settings. See e.g.,~\cite{Chatterjee2025quantum, Pace2025Tduality, Seiberg2025LSMCPT, seiberg2026torikleinbottlesmodulo,
fidkowski2025noninvertiblebosonicchiralsymmetry, thorngren2026chirallatticegaugetheories, Seiberg2024Majorana, Gioia2026exact, berkowitz2024exactlatticechiralsymmetry,
Kim2026fermi, Pace2025parity, liu2026anomaliesquantumspinsystems, lu2026fermionicvillainmodelexact, lewsmith2026infiniteorderlatticechiralanomalies,dharanikota202611dlatticediracfermions} for recent development of how fermionic anomalies are realized on lattice models.

In this work, we develop lattice characterization for finite internal fermionic ’t Hooft anomalies of exact symmetries in one and two spatial dimensions. We consider lattice systems whose local Hilbert space is a tensor product of a fermionic Fock space and a bosonic qudit Hilbert space. The fermionic symmetry is a finite group with the central extension $\mathbb{Z}_2^F \rightarrow G_f \rightarrow G_b$ characterized by an extension class $\omega_2\in H^2(BG_b,\mathbb{Z}_2)$. 
Such exact symmetry operators are generated by finite-depth fermionic quantum circuits on the lattice. 

In (1+1) dimensions, we associate to an exact finite internal $G_f$ symmetry a pair of anomaly indices
\begin{equation}
(n_2,\nu_3),
\qquad
n_2\in Z^2(BG_b,\mathbb{Z}_2),
\qquad
\nu_3\in C^3(BG_b,\mathbb{R}/\mathbb{Z}),
\end{equation}
satisfying
\begin{equation}
\delta\nu_3
=
\frac{1}{2}
\left(
n_2\cup n_2+\omega_2\cup n_2
\right)
\quad \mathrm{mod}\ 1.
\end{equation}
The index $n_2$ was originally introduced by Else and Nayak \cite{else2014}, measuring the fermion parity carried by the fusion of two symmetry defects. The index $\nu_3$ measures the bosonic part of the 't Hooft anomaly. The above consistency equations between $n_2, \nu_3$ reproduces the cohomological data for (2+1)D fermionic SPT phase that describes the inflow~\cite{gu2014, WangGu2020, williamson2017fermionicmatrixproductoperators, barkeshli2021invertible}.
We note that  \cite{okada2025anomaliesfermionicunitaryoperators} also obtains such operator-based anomaly characterization of $G_f$ symmetry in (1+1)D continuum QFT.
We determine gauge equivalences of these cochains and derive their stacking rule, thereby completing the description of fermionic anomaly index in (1+1)D.

For a trivial extension,
\begin{equation}
G_f=G_b\times\mathbb{Z}_2^F,
\end{equation}
we further show that an exact symmetry with trivial anomaly index $(n_2,\nu_3)=(0,0)$ is onsiteable, i.e., is transformed into onsite form after adding bosonic ancillary degrees of freedom and conjugating by a finite-depth circuit \cite{seifnashri2025disentangling}. 
The symmetry is consequently gaugeable and admits a symmetric SRE state. Thus, for $G_b\times\mathbb{Z}_2^F$, the pair $(n_2,\nu_3)$ gives a faithful characterization of the ’t Hooft anomaly of an exact lattice symmetry.

However, this lattice classification does not exhaust the anomalies allowed in continuum quantum field theory. The corresponding fermionic QFT anomaly contains an additional layer and is characterized schematically by~\cite{WangGu2020, barkeshli2021invertible}
\begin{equation}
(n_1,n_2,\nu_3),
\qquad
[n_1]\in H^1(BG_b,\mathbb{Z}_2).
\end{equation}
Because every exact lattice symmetry with trivial $(n_2,\nu_3)$ is onsiteable, a continuum anomaly whose only nontrivial component is $n_1$ cannot be realized by an exact internal symmetry on the lattice. More generally, continuum anomaly classes with nontrivial $n_1$ lie outside the image of the exact lattice anomaly index.

A simple example is furnished by $G_b=\mathbb{Z}_2$. In continuum (1+1)D fermionic QFT, the $\mathbb{Z}_2$ anomaly is classified by $\mathbb{Z}_8$~\cite{FidkowskiKitaev1d, Kapustin2015spinbordism}. By contrast, the exact lattice $\mathbb{Z}_2$ symmetry described by $(n_2,\nu_3)$ realizes only the even elements, which form a $\mathbb{Z}_4$ subgroup of the full $\mathbb{Z}_8$ classification. 
In particular, four copies of any exact lattice $\mathbb{Z}_2$ symmetry have trivial lattice anomaly and are onsiteable. The odd mod-$8$ anomaly therefore cannot be implemented by an exact $\mathbb{Z}_2$ symmetry on a tensor product Hilbert space.\footnote{This no-go statement is reminiscent of recent developments concerning which non-invertible symmetries can be realized as exact fusion category symmetries on the lattice~\cite{evans2026operatoralgebraicapproachfusion, 
inamura2026remarksnoninvertiblesymmetriestensor,wen2026noninvertiblesymmetriestensorproducthilbert}. It is known that a fusion category symmetry can be realized as an exact symmetry of a tensor-product Hilbert space if and only if the unitary fusion category is integral~\cite{evans2026operatoralgebraicapproachfusion, 
inamura2026remarksnoninvertiblesymmetriestensor}. This result can be established by characterizing the dimensions of defect Hilbert spaces on the lattice, dubbed the lattice quantum dimension.
Since the anomalous $\mathbb{Z}_2$ symmetry can be understood in terms of a fusion supercategory \cite{Bhardwaj2025fermionic}, we expect our no-go statement to follow as a corollary of a more general statement: a fermionic fusion supercategory symmetry can be realized as an exact symmetry if and only if the category is integral. We emphasize, however, that our proof makes no assumption about the lattice quantum dimension associated with the anomalous $\mathbb{Z}_2$ symmetry and instead provides a direct proof based on onsiteability.}
Known lattice realizations of these odd anomaly classes instead involve an emanant symmetry implementing Majorana translation~\cite{Seiberg2024Majorana, Aksoy2021LSM}. 
We note that  \cite{czajka2025anomalieslatticehomotopyquantum} also discusses the lattice mod 4 anomaly of $\Z_2$ symmetry defined through blend equivalence of quantum cellular automata (QCA).
Our approach instead characterizes the anomaly directly by its dynamical consequence: a symmetry is anomaly-free precisely when it admits a symmetric SRE state.
We then establish the mod 4 anomaly by proving four copies of any exact $\Z_2$ symmetry are onsiteable.

In (2+1) dimensions, we formulate a hierarchical set of three fermionic anomaly layers,
\begin{equation}
n_2,~n_3,~\nu_4~,
\end{equation}
for generic finite internal symmetry $\mathbb{Z}_2^F \rightarrow G_f \rightarrow G_b$. These indices correspond to the cohomological data for (3+1)D $G_f$ SPT phase that describes the inflow~\cite{WangGu2020, Delmastro2021global, Barkeshli:2022edm}.
The first index is dubbed the Majorana layer, represented by
\begin{equation}
[n_2]\in H^2(BG_b,\mathbb{Z}_2)~.
\end{equation}
It is detected by truncating the symmetry to a disk and examining the 1D fermionic QCA left with the projective action of the symmetry: $n_2(g,h)$ distinguishes an ordinary rational QCA index~\cite{GNVW2012, Po2016chiral} from an intrinsically fermionic index containing a factor of $\sqrt{2}$, corresponding to Majorana translation~\cite{Fidkowski2019fermionicQCA, trezzini2025fermioniccellularautomatadimension}.

After the Majorana layer has been trivialized, the boundary QCA can be reduced to a finite-depth circuit, up to a non-anomalous bosonic QCA index. Truncating this boundary circuit to an interval produces endpoint associator operators whose fermion parity defines the complex-fermion layer
\begin{equation}
[n_3]\in H^3(BG_b,\mathbb{Z}_2)~.
\end{equation}
When this layer is also trivialized, the remaining endpoint associativity relation is satisfied up to a $U(1)$ phase, defining the bosonic layer
\begin{equation}
[\nu_4]\in H^4(BG_b, \R/\Z)~.
\end{equation}
Thus, the three indices arise successively from Majorana transport, fermion parity, and a purely bosonic part of 't Hooft anomaly.

We establish the dynamical consequence of each anomaly layer. A nontrivial $n_2$, $n_3$, or $\nu_4$ obstructs the existence of a symmetric SRE state. When the extension class is trivial, so that
\begin{equation}
G_f=G_b\times\mathbb{Z}_2^F~,
\end{equation}
the obstruction is stronger: a nontrivial anomaly index forbids any symmetric invertible state.

Importantly, the anomaly index attached to an exact lattice symmetry can retain information that the continuum anomaly classification discards, and this residual information has dynamical consequences. We illustrate this with an explicit $\mathbb{Z}_4^F$ symmetry in (2+1) dimensions, generated by $U_g$ with $U_g^2 = (-1)^F$, which carries the nontrivial Majorana layer index $n_2 = \omega_2$. We show that this index forbids every symmetric invertible state whose chiral central charge is an integer $c_- \in \mathbb{Z}$. Since invertible phases preserved by an \textit{onsite} $\mathbb{Z}_4^F$ symmetry are precisely those with $c_- \in\mathbb{Z}$~\cite{barkeshli2021invertible, kobayashi2026remarks}, this lattice symmetry is incompatible with every invertible phase that an onsite realization can support.

The symmetry does not, however, forbid symmetric invertible phases altogether. Ref.~\cite{kobayashi2026remarks} recently showed that a lattice $\mathbb{Z}_4^F$ symmetry carrying the same index $n_2=\omega_2$ is realized exactly in the $p+ip$ superconductor, though by an exponentially quasi-local operator rather than a strictly finite-depth circuit.\footnote{\label{fn:quasilocal}The symmetry operator of Ref.~\cite{kobayashi2026remarks} is obtained by spectral flattening of a local gapped BdG Hamiltonian, so its action on local operators has exponentially decaying spatial tails, while satisfying the $\mathbb{Z}_4^F$ algebra exactly. We expect the anomaly indices to remain well defined in this setting, since the classification of one-dimensional (F)QCAs is robust against such tails~\cite{ranard2026approximateqcasdimensionusing, Ranard2022converse}.} Stacking with onsite $\mathbb{Z}_4^F$-symmetric invertible phases then gives symmetric invertible states with $c_- \in \mathbb{Z}+{1}/{2}$. The lattice anomaly index therefore does not eliminate the symmetry-compatible invertible phases; it shifts the allowed chiral central charges from $\mathbb{Z}$ to $\mathbb{Z}+{1}/{2}$.

In the continuum, the anomaly data $(n_2,n_3,\nu_4)$ is defined modulo the equivalence $n_2\sim n_2+\omega_2$, which is implemented physically by stacking a $p+ip$ state and hence shifts $c_-$ by $1/2$; the class $n_2=\omega_2$ is therefore trivial there~\cite{WangGu2020}. Indeed, the continuum 't Hooft anomaly of $\Z_4^F$ symmetry has a trivial classification in (2+1)D~\cite{Garcia_Etxebarria_2019}. Since this identification relates invertible phases with $c_-\in\mathbb{Z}$ to those with $c_-\in\mathbb{Z}+{1}/{2}$, the continuum classification of anomalies cannot record which of the two families a given symmetry permits. A fixed lattice symmetry operator does: it selects one of them. The lattice index thus acts not as an obstruction to symmetric invertible phases but as a selection rule on which invertible states are compatible with the given microscopic symmetry action, and the phases it selects are precisely those that require a non-onsite realization of $\mathbb{Z}_4^F$.

More generally, this example indicates that the map from lattice anomalies to continuum '{}t~Hooft anomalies has a nontrivial kernel: an exact lattice symmetry can obstruct symmetric SRE states while descending to a trivial anomaly in the continuum. We return to this in Sec.~\ref{sec:discussion}.

This example demonstrates that a lattice obstruction to symmetric SRE phases need not correspond to a nontrivial continuum QFT anomaly. An exact lattice symmetry may exclude conventional symmetric SRE phases even when its continuum ’t Hooft anomaly is trivial. More generally, such a lattice anomaly can modify the set of symmetry-compatible invertible phases, permitting phases that require a non-onsite realization of the symmetry.

Our result motivates broader problems of which continuum fermionic anomalies admit  microscopic lattice realizations by an exact symmetry, and which lattice obstructions to SRE states have no counterpart in the continuum anomaly classification.
We close this paper with several conjectures regarding these questions.

This paper is organized as follows. In Sec.~\ref{sec:else nayak in 1D} we describe anomaly indices in (1+1)D lattice systems.
In Sec.~\ref{sec:onsiteability} we establish the onsiteability of symmetries with trivial fermionic anomaly indices in (1+1)D. In Sec.~\ref{sec:emanant} we show that the $[n_1]\in H^1(BG_b,\Z_2)$ layer of the anomaly in (1+1)D cannot be realized by an exact symmetry on the lattice. In Sec.~\ref{sec:anomaly in 2+1D} we construct a hierarchy of anomaly indices in (2+1)D. In Sec.~\ref{sec:dynamical} we establish the dynamical consequence of 't Hooft anomalies in (2+1)D. In Sec.~\ref{sec:anomaly free sym forbids SRE} we construct a lattice $\Z_4^F$ symmetry that forbids invertible states while carrying trivial QFT anomaly in continuum. In Sec.~\ref{sec:discussion} we conclude this paper with discussions and conjectures on anomalies accessible to exact symmetries on lattices.

\section{Else-Nayak index for fermionic anomalies in (1+1)D}
\label{sec:else nayak in 1D}
Let us consider a 1D chain formed by a complex fermion $\{c_j, c^\dagger_j\}$ and onsite (bosonic) Hilbert space $\mathcal{H}^{(b)}_j$ at each site $j$. That is, onsite Hilbert space is a tensor product $\mathcal{H}^{(b)}_j\otimes \mathcal{H}^{(f)}_j$, where $\mathcal{H}^{(f)}_j$ is a Fock space of a single complex fermion. Here we included additional bosonic Hilbert space $\{\mathcal{H}^{(b)}_j\}$ to study generic 1D systems that involves both fermions and qudits.

We consider finite internal symmetry $G_f$ of the chain. Any fermionic system possesses fermion parity symmetry $\Z_2^F\subset G_f$ generated by $(-1)^F := \prod_j (-1)^{c^\dagger_j c_j}$. The symmetry generally has the form of central extension: $\Z_2^F\to G_f\to G_b$, with the extension class characterized by $\omega_2\in Z^2(BG_b, \Z_2)$. Since the symmetry is finite and internal in 1D, the symmetry is generated by a finite depth quantum circuit (FDQC). Each local circuit of FDQC is fermion parity even. The FDQC generating the symmetry is denoted by $U(g)$ with $g\in G_b$, satisfying the group algebra
\begin{equation}
U(g)U(h) = [(-1)^{F}]^{\omega_2(g,h)}U(gh)~, 
\end{equation}
and commuting with fermion parity
\begin{align}
    U(g) (-1)^F = (-1)^F U(g)~.
    \label{eq:commutativity of U(g) with fermion parity}
\end{align}

The 't Hooft anomaly of this $G_f$ symmetry in 1D has been essentially described in \cite{else2014, Seifnashri2024lsm}, but here let us provide detailed characterization of the 't Hooft anomaly for completeness.
The first step is to take any truncation of the FDQC $U(g)$ within an interval $I$, denoted by $U_I(g)$. The truncated operators satisfy
\begin{equation}
U_I(g)U_I(h) = \Omega_{\partial I}(g,h)[(-1)^{F_I}]^{\omega_2(g,h)}U_I(gh)~,
\end{equation}
where $(-1)^{F_I}:= \prod_{j\in I} (-1)^{c^\dagger_j c_j}$, and $\Omega_{\partial I}(g,h)$ is an operator localized at the ends of the interval $\partial I$. Writing two ends of $\partial I$ as $l, r$, $\Omega_{\partial I}(g,h)$ has the form of
\begin{align}
    \Omega_{\partial I}(g,h) = \Omega_l(g,h)\Omega_r(g,h)~,
\end{align}
where $\Omega_l, \Omega_r$ are local operators at left and right ends respectively. 

The first anomaly index is then given by the fermion parity of the local operator $\Omega_l(g,h)$:
\begin{align}
    n_2(g,h) = \begin{cases}
0, &  \text{$\Omega_l(g,h)$ is fermion parity even},\\
1, & \text{$\Omega_l(g,h)$ is fermion parity odd}.
\end{cases}
\end{align}
This 2-cochain $n_2$ is the same for the left and right end, so we do not distinguish $(n_2)_l$ and $(n_2)_r$ and simply write them as $n_2$.

The operator $\Omega_{\partial I}$ then satisfies the cocycle condition
\begin{align}
    \Omega_{\partial I}(g,h) [(-1)^{F_I}]^{\omega_2(g,h)} \Omega_{\partial I}(gh,k)[(-1)^{F_I}]^{\omega_2(gh,k)} = \ ^g\left[\Omega_{\partial I}(h,k) [(-1)^{F_I}]^{\omega_2(h,k)}\right]\Omega_{\partial I}(g,hk) [(-1)^{F_I}]^{\omega_2(g,hk)}
\end{align}
where $^g\mathcal{O}:= U_I(g) \mathcal{O} U_I(g)^\dagger$.  $\{\Omega_{\partial I}\}$ are fermion parity even and commute with $(-1)^{F_I}$, hence $\Omega_l$ satisfies the cocycle condition

\begin{align}
    \Omega_{l}(g,h) 
    \Omega_{l}(gh,k)
    = \ ^g\left[\Omega_{l}(h,k) 
    \right]\Omega_{l}(g,hk) 
    \cdot e^{2\pi i \nu_3(g,h,k)}~,
    \label{eq:def of nu3}
\end{align}
where $e^{2\pi i \nu_3(g,h,k)}$ is a phase factor which gives the second anomaly index. This is regarded as a 3-cochain $\nu_3\in C^3(BG_b,\mathbb{R}/\mathbb{Z})$.
Comparing fermion parity of two sides of the above equation, we get $\delta n_2 =0$, therefore $n_2\in Z^2(BG_b,\Z_2)$.

The anomaly indices $n_2, \nu_3$ are subject to the following equation:
\begin{align}
    \delta \nu_3 = \frac{1}{2}\left(n_2\cup n_2 + \omega_2\cup n_2\right) \quad \text{mod 1}~.
    \label{eq:guwen in 1+1D}
\end{align}
This is derived by evaluating $\Omega_{l}(g,h) 
\Omega_{l}(gh,k)
\Omega_{l}(ghk,l)
$ in the following two ways:
\begin{align}
    \begin{split}
        &\Omega_{l}(g,h)
        \Omega_{l}(gh,k)
        \Omega_{l}(ghk,l) 
        \\
        =& e^{2\pi i \nu_3(g,h,k)} \times {^g}\left[\Omega_{l}(h,k) 
        \right]
        \Omega_{l}(g,hk) 
        \Omega_{l}(ghk,l)
        \\
         =& e^{2\pi i \nu_3(g,h,k)}e^{2\pi i \nu_3(g,hk,l)} \times {^g}\left[\Omega_{l}(h,k)
         \right] ^{g}\left[\Omega_{l}(hk,l) 
         \right]\Omega_{l}(g,hkl) 
         \\
         =& e^{2\pi i \nu_3(g,h,k)}e^{2\pi i \nu_3(g,hk,l)}e^{2\pi i \nu_3(h,k,l)} \times {^g}\left[{^{h}}\left[\Omega_{l}(k,l) 
         \right] \right] ^{g}\left[\Omega_{l}(h,kl) 
         \right]\Omega_{l}(g,hkl)~, 
    \end{split}
\end{align}
and
\begin{align}
    \begin{split}
        &\Omega_{l}(g,h) 
        \Omega_{l}(gh,k)
        \Omega_{l}(ghk,l)
        \\
        =& e^{2\pi i \nu_3(gh,k,l)}\times \Omega_{l}(g,h) 
        \cdot {^{gh}}\left[\Omega_{l}(k,l) 
        \right]\Omega_{l}(gh,kl) 
        \\
        =& e^{2\pi i \nu_3(gh,k,l)}(-1)^{n_2(g,h)n_2(k,l)+\omega_2(g,h)n_2(k,l) } \times {^g}\left[{^{h}}\left[\Omega_{l}(k,l) 
        \right] \right]\Omega_{l}(g,h) 
        \Omega_{l}(gh,kl) 
        \\
        =& e^{2\pi i \nu_3(gh,k,l)}e^{2\pi i \nu_3(g,h,kl)}(-1)^{n_2(g,h)n_2(k,l)+\omega_2(g,h)n_2(k,l)} \times {^g}\left[{^{h}}\left[\Omega_{l}(k,l) 
        \right] \right] ^{g}\left[\Omega_{l}(h,kl) 
        \right]
        \Omega_{l}(g,hkl)~. 
    \end{split}
\end{align}

\subsection{Invariance of anomaly indices}

There are a number of ambiguities to redefine the anomaly index $(n_2, \nu_3)$. First, taking different circuit decomposition of a given symmetry operator $U(g)$ generally yields a different choice of the truncation $U_I(g)$ and shifts $\Omega_{\partial I}$.\footnote{For instance, the identity operator could be expressed by an FDQC $\propto \prod_{j\in\text{odd}} \gamma_j\gamma_{j+1} \prod_{j\in\text{even}} \gamma_j\gamma_{j+1}$ using a Majorana fermion $\gamma_j:=(c_j+c^\dagger_j)/\sqrt{2}$, then its truncation to an interval has the form of $\gamma_l\gamma_r$ and shifts $\Omega_l$ by multiplication of $\gamma_l$.} Such ambiguity generally shifts $U_I(g)$ by $U'_I(g)=\mathcal{O}_l(g)\mathcal{O}_r(g)U_I(g)$, where $\mathcal{O}_l, \mathcal{O}_r$ are local operators at the left, right ends carrying the same fermion parity. Writing the fermion parity of $\mathcal{O}_l(g)$ by a 1-cochain $\chi(g)\in C^1(BG_b,\Z_2)$, this shifts the anomaly index $n_2(g,h)$ by a coboundary
\begin{align}
    n'_2 =  n_2 + \delta\chi~.
\end{align}

Let us then study how the index $\nu_3$ gets shifted.
After the redefinition, $\Omega_l$ becomes
\begin{align}
    \Omega'_l(g,h) =\mathcal{O}_l(g)\ ^g\mathcal{O}_l(h) \ \Omega_{l}(g,h)\ \mathcal{O}_l(gh)^{-1}~.
\end{align}
 Plugging this expression to \eqref{eq:def of nu3}, a tedious calculation shows that $\nu_3$ is shifted by
\begin{align}
    \nu'_3= \nu_3+ \frac{1}{2}(\chi\cup n_2 + n_2\cup \chi + \chi\cup \delta\chi + \omega_2\cup \chi)~.
\end{align}
Note that the new anomaly indices $(n'_2, \nu'_3)$ again satisfy the equation \eqref{eq:guwen in 1+1D}.

On top of the above redefinition, one can always redefine $\Omega_l(g,h)$ by a phase factor $e^{2\pi i\lambda(g,h)}$ using $\lambda\in C^2(BG_b, \R/\Z)$, which shifts the index by $(n_2,\nu_3)\to (n_2, \nu_3+\delta\lambda)$. Summarizing, the anomaly index is invariant under possible ambiguities up to the following gauge transformation
\begin{align}
    (n_2, \nu_3)\to \left(n_2+\delta\chi,\nu_3+ \frac{1}{2}(\chi\cup n_2 + n_2\cup \chi + \chi\cup \delta\chi + \omega_2\cup \chi) + \delta\lambda \right)~, \quad \chi\in C^1(BG_b, \Z_2), \lambda\in C^2(BG_b, \R/\Z)~.
    \label{eq:gauge trans of gu wen data}
\end{align}

\subsection{Stacking rule}
Let us consider symmetry operator $U(g)\otimes U'(g)$ obtained by tensoring two symmetry operators $U(g), U'(g)$. Using the anomaly indices of $U, U'$ by $(n_2, \nu_3), (n'_2, \nu'_3)$, one can express the anomaly index of the tensored symmetry operator $U\otimes U'$. It is given by
\begin{align}
    \left(n_2+n'_2, \nu_3 + \nu'_3 + \frac{1}{2}(n_2\cup_1 n'_2)\right)~.
    \label{eq:stacking in 1D}
\end{align}
This formula is derived by plugging the expression $[\Omega_l]_{U\otimes U'} = \Omega_l\Omega_l'$ into \eqref{eq:def of nu3}, then the fermionic signs from commuting $\Omega_l, \Omega_l'$ gives the phase factor $(-1)^{n_2\cup_1 n'_2(g,h,k)}$ that shifts $e^{2\pi i \nu_3}$ as above.

\section{Onsiteability of anomaly-free symmetries in (1+1)D}
\label{sec:onsiteability}
We show that the $G_f$ symmetry in (1+1)D with the trivial anomaly index $(n_2,\nu_3)$ is onsiteable and gaugeable, therefore free of 't Hooft anomaly on the lattice. Here we assume that the extension class for $\Z_2^F\to G_f\to G_b$ is trivial; $\omega_2(g,h)=0$, so the symmetry has the form of $G_f=G_b\times \Z_2^F$.
The onsiteability of the $G_f$ symmetry $\{U(g)\}$ with the trivial anomaly index is established in exactly the same fashion as the bosonic case, which has been developed in \cite{seifnashri2025disentangling}; we will explicitly construct a finite depth circuit $\mathcal{W}$ that disentangles the $G_b$ symmetry $U(g)\otimes \mathcal{X}_g$ of the enlarged Hilbert space into an onsite form, where $\mathcal{X}_g$ is the bosonic onsite $G_b$ symmetry operator acting on the ancillary Hilbert space.

To construct the disentangler, we adopt the conventions of \cite{seifnashri2025disentangling} to describe the 't Hooft anomaly on the 1D lattice. We work on the infinite chain and consider a 1D mesoscopic lattice, formed by blocks of the original microscopic lattice with the constant length $l$. 

For each site $j$ of the mesoscopic lattice, we introduce a truncated symmetry operator $U^g_{\leq j}$ for $g\in G_b$ that acts on the left half-infinite line ending at the site $j$. 
Such a half-line truncation creates a symmetry defect when acting on a quantum state. The product $U^g_{\leq j-1}U^h_{\leq j}$ creates a $g$ defect and an $h$ defect at the adjacent sites, while $U^{gh}_{\leq j}$ creates the $gh$ defect that corresponds to fusing them. 
We therefore define the operator that corresponds to fusing the $g,h$ defects
\begin{equation}\label{eq:fusiondef}
 \lambda_j(g,h) \propto 
 U^{gh}_{\leq j} \left(U^g_{\leq j-1}U^h_{\leq j}\right)^{-1}.
\end{equation}
Note that this operator $\lambda_j(g,h)$ is supported at the edge $\langle j,j+1\rangle$ of the mesoscopic lattice. 
The mesoscopic block size is chosen large compared with the circuit depth and the range of the local circuits, so that $\lambda_{j}$ and $\lambda_{j+2}$ commute with each other.

We introduce the first anomaly index by the fermion parity of $\lambda_j(g,h)$:
\begin{align}
    n_2(g,h) = \begin{cases}
0, &  \text{$\lambda_j(g,h)$ is fermion parity even},\\
1, & \text{$\lambda_j(g,h)$ is fermion parity odd}.
\end{cases}
\end{align}
We note that this $n_2$ index has been introduced in \cite{Seifnashri2024lsm}.
This cochain $n_2$ is independent of the choice of the site $j$; moving the truncation $U^g_{\leq j}$ to $U^g_{\leq j-1}$ is performed by acting a finite depth circuit with even fermion parity, therefore the operators $\lambda_j(g,h) \propto 
U^{gh}_{\leq j} \left(U^g_{\leq j-1}U^h_{\leq j}\right)^{-1}$ and $\lambda_{j-1}(g,h) \propto  
U^{gh}_{\leq j-1} \left(U^g_{\leq j-2}U^h_{\leq j-1}\right)^{-1}$ carry the same fermion parity. 
The cochain $n_2$ also satisfies
\begin{align}
    n_2(g,1) = n_2(1,g) = 0~.
\end{align}

The fusion operators $\lambda_j(g,h)$ then satisfy the equation
\begin{equation} \label{eq:Fmove}
 \lambda_j(g,hk)\lambda_{j-1}(g,1)\lambda_j(h,k) = F_j(g,h,k)\lambda_j(gh,k)\lambda_{j-1}(g,h)~,
\end{equation}
where $F_j(g,h,k)$ is a phase factor that is used later to define the second anomaly index. Comparing the fermion parity of both sides in the this equation, we see that $n_2$ is closed: $\delta n_2= 0$. 

The cochains $n_2, \{F_j\}$ are subject to the equation
\begin{align}
    \frac{F_{j}(g_1, g_2, g_3)F_{j+1}(g_1, g_2g_3, g_4) F_{j+1}(g_2, g_3, g_4)}{F_{j+1}(g_1g_2,g_3,g_4) F_{j+1}(g_1, g_2, g_3g_4) } = (-1)^{n_2(g_1,g_2)n_2(g_3,g_4)} \times F_{j}(g_1, g_2, 1)~.
    \label{eq:modified pentagon in 1+1D}
\end{align}

This is derived by evaluating the same product of $\lambda$ operators in two different ways. On one hand,
\begin{align}
\begin{split}
    &\lambda_{j+1}(g_1, g_2 g_3 g_4)  \lambda_{j}(g_1, 1) \lambda_{j-1}(g_1, 1) \lambda_{j+1}(g_2, g_3 g_4) \lambda_j(g_2, 1) \lambda_{j+1}(g_3, g_4) \\
    =& \lambda_{j+1}(g_1, g_2 g_3 g_4)  \lambda_{j}(g_1, 1) \lambda_{j+1}(g_2, g_3 g_4) \lambda_{j-1}(g_1, 1)  \lambda_j(g_2, 1) \lambda_{j+1}(g_3, g_4) \\
    =& F_{j+1}(g_1, g_2, g_3g_4)\times  \lambda_{j+1}(g_1g_2, g_3g_4) \lambda_j(g_1, g_2) \lambda_{j-1}(g_1, 1)  \lambda_j(g_2, 1) \lambda_{j+1}(g_3, g_4) \\
    =& F_{j+1}(g_1, g_2, g_3g_4) F_{j}(g_1, g_2, 1)\times \lambda_{j+1}(g_1g_2, g_3g_4) \lambda_j(g_1g_2,1) \lambda_{j-1}(g_1,g_2) \lambda_{j+1}(g_3, g_4) \\
    =&  (-1)^{n_2(g_1,g_2)n_2(g_3,g_4)} F_{j+1}(g_1, g_2, g_3g_4) F_{j}(g_1, g_2, 1)\times \lambda_{j+1}(g_1g_2, g_3g_4) \lambda_j(g_1g_2,1) \lambda_{j+1}(g_3, g_4) \lambda_{j-1}(g_1,g_2)  \\
    =&  (-1)^{n_2(g_1,g_2)n_2(g_3,g_4)} F_{j+1}(g_1g_2,g_3,g_4) F_{j+1}(g_1, g_2, g_3g_4) F_{j}(g_1, g_2, 1)\times \lambda_{j+1}(g_1g_2g_3,g_4) \lambda_j(g_1g_2,g_3) \lambda_{j-1}(g_1,g_2)~.  \\
    \end{split}
\end{align}
On the other hand,
\begin{align}
    \begin{split}
        &\lambda_{j+1}(g_1, g_2 g_3 g_4)  \lambda_{j}(g_1, 1) \lambda_{j-1}(g_1, 1) \lambda_{j+1}(g_2, g_3 g_4) \lambda_j(g_2, 1) \lambda_{j+1}(g_3, g_4) \\
=& F_{j+1}(g_2, g_3, g_4)\times \lambda_{j+1}(g_1, g_2 g_3 g_4)  \lambda_{j}(g_1, 1) \lambda_{j-1}(g_1, 1) \lambda_{j+1}(g_2g_3, g_4) \lambda_j(g_2, g_3) \\
=& F_{j+1}(g_2, g_3, g_4)\times \lambda_{j+1}(g_1, g_2 g_3 g_4)  \lambda_{j}(g_1, 1)  \lambda_{j+1}(g_2g_3, g_4) \lambda_{j-1}(g_1, 1) \lambda_j(g_2, g_3) \\
=& F_{j+1}(g_1, g_2g_3, g_4) F_{j+1}(g_2, g_3, g_4)\times \lambda_{j+1}(g_1 g_2 g_3,  g_4) \lambda_j(g_1, g_2g_3) \lambda_{j-1}(g_1, 1) \lambda_j(g_2, g_3) \\
=& F_{j}(g_1, g_2, g_3)F_{j+1}(g_1, g_2g_3, g_4) F_{j+1}(g_2, g_3, g_4)\times \lambda_{j+1}(g_1 g_2 g_3,  g_4)  \lambda_{j}(g_1 g_2, g_3)   \lambda_{j-1}(g_1, g_2)~. \\
    \end{split}
\end{align}
The second anomaly index is now defined as
\begin{align}
    e^{2\pi i \nu_j(g,h,k)} = \frac{F_j(g,h,k)}{F_j(g,h,1)}~.
\end{align}
Using \eqref{eq:modified pentagon in 1+1D} for generic $g_1,g_2, g_3, g_4$ and the one with $g_4=1$, we obtain
\begin{align}
    \delta \nu_j=\frac{1}{2}n_2\cup n_2~, \quad \delta\alpha_j = \nu_j-\nu_{j-1}~\quad \mod 1~,
\end{align}
where the first equation reproduces \eqref{eq:guwen in 1+1D} with $\omega_2=0.$ In the above, we have introduced a 2-cochain $\alpha_j$ by
\begin{align}
    e^{2\pi i \alpha_j(g,h)} = F_j(g,h,1)~.
\end{align}
Since $\nu_j=\nu_{j-1}+\delta\alpha_j$, the equivalence class of anomaly indices up to the gauge transformations \eqref{eq:gauge trans of gu wen data} is independent of the labeling $j$, therefore we simply write $\nu_j=\nu_3$ using some fixed $j$. The above $(n_2, \nu_3)$ gives a characterization of the 't Hooft anomaly equivalent to the Else-Nayak index introduced in Sec.~\ref{sec:else nayak in 1D}.

When the anomaly index is trivial, $(n_2,\nu_3)=(0,0)$ in the above description, one can use the exactly same disentangler $\mathcal{W}$ as \cite{seifnashri2025disentangling} to transform the symmetry  
into onsite form by coupling to the ancilla and conjugation action of $\mathcal{W}$. 
We first couple the original Hilbert space with the ancillary Hilbert space $\mathcal{H}'=\bigotimes_j \mathbb{C}^{|G_b|}$, i.e., onsite Hilbert space has $|G_b|$ dimensions and introduced at each block of the mesoscopic lattice.
We then consider the $G_b$ symmetry of the enlarged Hilbert space $U(g)\otimes \mathcal{X}_g$, where $\mathcal{X}_g$ is onsite $G_b$ symmetry acting on the ancillary Hilbert space with the form of
\begin{align}
    \mathcal{X}_g = \bigotimes_j L_j^{g}, \quad L_j^{g} = \sum_{s_j\in G_b}\ket{gs_j}\bra{s_j}~,
\end{align}
where we labeled basis states of the onsite Hilbert space by $\{\ket{s}\}$ using group elements $s\in G_b$. 

When $(n_2,\nu_3)$ indices are trivial, one can choose the fusion operators $\{\lambda_j\}$ such that they are fermion parity even, and the $F$ symbols $\{F_j\}$ are all trivial. Then, the following FDQC $\mathcal{W}$ is shown to disentangle the symmetry operator $U(g)\otimes \mathcal{X}_g$ \cite{seifnashri2025disentangling}:
\begin{equation}\label{eq:Wblock}
\mathcal W=\sum_{\{s_j\}}\Lambda_{\{s_j\}}\otimes
|\{s_j\}\rangle\langle\{s_j\}|~, \quad \Lambda_{\{s_j\}}=\prod_j\lambda_j\bigl(s_j,\;\overline{s}_js_{j+1}\bigr)~,  
\end{equation}
with ordering of the product $\cdots\lambda_{j+1}\lambda_j\lambda_{j-1}\cdots$. Namely, $\mathcal{W}$ satisfies
\begin{equation}\label{eq:disentangle}
\mathcal W^{\dagger}\,\bigl(U(g)\otimes\mathcal X^g\bigr)\,\mathcal W
\;=\;
\mathbbm 1 \otimes\mathcal X^g~.
\end{equation}
This implies that the $G_b$ symmetry is onsiteable when $\{F_j\}$ are all trivial, which is the case with the trivial anomaly index $(n_2,\nu_3)=(0,0)$.

\section{$\Z_2$ symmetry with odd mod-8 anomaly cannot be exact}
\label{sec:emanant}
We have established that the $G_f=G_b\times\Z^F_2$ symmetry with the trivial anomaly index $(n_2,\nu_3)=(0,0)$ is onsiteable, therefore admits a symmetric SRE state and anomaly-free. This implies that $(n_2,\nu_3)$ index gives a faithful characterization of 't Hooft anomaly of exact $G_b\times \Z_2^F$ symmetry on the lattice.

The situation is different in (1+1)D continuum QFT. There, the ’t Hooft anomaly of $G_f$ contains an additional layer, $[n_1]\in H^1(BG_b,\Z_2)$, and characterized by the triple $(n_1, n_2, \nu_3)$. In particular, when $G_f=G_b\times \Z_2^F$, the triple $(n_1,0,0)$ with generic $[n_1]\in H^1(BG_b,\Z_2)$ defines a nontrivial 't Hooft anomaly. 
This continuum anomaly, however, cannot be realized by an exact lattice symmetry. Indeed, any exact lattice symmetry with $n_2=0$ and $\nu_3=0$ is onsiteable and hence anomaly-free. Therefore, an anomaly whose only nontrivial component is $n_1$ is invisible to the lattice anomaly index and cannot occur for an exact symmetry on the lattice. More generally, any continuum anomaly $(n_1,n_2,\nu_3)$ with nontrivial $[n_1]$ lies outside the set of anomalies realizable by exact lattice symmetries.

Let us see this directly when $G_b=\Z_2$. While its continuum QFT anomaly is classified by $\Z_8$, the lattice exact symmetry $(n_2,\nu_3)$ only realizes the $\Z_4$ subgroup of the whole QFT anomaly. In particular, let us consider any $\Z_2$ symmetry generator $U(g)$ and take four copies of them, $U(g)\otimes U(g)\otimes U(g)\otimes U(g)$. By repeated use of the stacking rule presented in \eqref{eq:stacking in 1D}, the symmetry $U\otimes U\otimes U\otimes U$ carries the trivial lattice anomaly index, $(n_2,\nu_3)_{U\otimes U\otimes U\otimes U} = (0,0)$, hence onsiteable.\footnote{To see this explicitly, first take two copies $U\otimes U$. The stacking rule gives $n_2^{(2)}=n_2+n_2=0$; equivalently, the doubled fusion operators $\lambda^{(2)}=\lambda\otimes\lambda$ can be chosen fermion parity even. Further stacking four copies $(U\otimes U)\otimes(U\otimes U)$ then gives $[\nu_3^{(4)}]=2[\nu_3^{(2)}]=0$ since  $[\nu_3^{(2)}] \in H^3(B\mathbb Z_2,U(1))$ when $n_2^{(2)}=0$ and $H^3(B\mathbb Z_2,U(1))=\Z_2$. } 
Indeed, known realizations of the odd element of the mod-8 anomaly in lattice models is by an emanant symmetry that generates Majorana translation, instead of exact $\Z_2$ symmetry~\cite{Seiberg2024Majorana, Aksoy2021LSM}.

\section{Index for fermionic anomalies in (2+1)D}
\label{sec:anomaly in 2+1D}
Let us consider the finite 0-form internal symmetry in (2+1)D with the group structure
$\Z_2^F\to G_f\to G_b$, with the extension class characterized by $\omega_2\in Z^2(BG_b, \Z_2)$. Following the setup of 1D case, we again consider the Hilbert space formed by onsite Hilbert space $\mathcal{H}^{(b)}_j\otimes \mathcal{H}^{(f)}_j$ at each site $j$. Below we describe the 't Hooft anomaly indices of the symmetry.

\subsection{Majorana layer}
The symmetry operator satisfies the group algebra,
\begin{equation}
U(g)U(h) = [(-1)^{F}]^{\omega_2(g,h)}U(gh)~.
\end{equation}
Since the fermionic QCA in 2D space is trivial up to circuit multiplications~\cite{kwan2026fermionicQCA}, the symmetry operator $U(g)$ is FDQC. Let us first truncate the symmetry $U(g)$ within a disk region $A$, and define the 1D operator $\Omega_{\partial A}(g,h)$ as
\begin{equation}
U_A(g)U_A(h) = \Omega_{\partial A}(g,h)[(-1)^{F_A}]^{\omega_2(g,h)}U_A(gh)~,
\label{eq:projective UA}
\end{equation}
where $\Omega_{\partial A}(g,h)$ is a 1D QCA supported in the vicinity of the ring $\partial A$. 

While a 1D bosonic QCA the GNVW index takes values in the positive rational numbers $\Q_+$~\cite{GNVW2012}, 1D fermionic QCAs admit an additional, intrinsically fermionic possibility~\cite{Fidkowski2019fermionicQCA}: the translation of a single Majorana fermion, then the index is valued in
\begin{equation}
\indF(\Omega_{\partial A})\in \mathbb{Q}_+\cup \sqrt{2} \mathbb{Q}_+~,
\end{equation}
where the $\sqrt{2}$ index corresponds to the Majorana translation. This allows us to define the $\Z_2$-valued cochain by
\begin{align}
    n_2(g,h) = \begin{cases}
0, &  \text{$\text{Ind}_F(\Omega_{\partial A}(g,h)) \in \mathbb{Q}_+$}~,\\
1, & \text{$\text{Ind}_F(\Omega_{\partial A}(g,h)) \in \sqrt{2}\mathbb{Q}_+$}~.
\end{cases}
\end{align}
which defines the first anomaly index. By evaluating $U_A(g)U_A(h)U_A(k)$ in two ways, we obtain
\begin{align}
    \Omega_{\partial A}(g,h) [(-1)^{F_A}]^{\omega_2(g,h)} \Omega_{\partial A}(gh,k)[(-1)^{F_A}]^{\omega_2(gh,k)} = \ ^g\left[\Omega_{\partial A}(h,k) [(-1)^{F_A}]^{\omega_2(h,k)}\right]\Omega_{\partial A}(g,hk) [(-1)^{F_A}]^{\omega_2(g,hk)}
\end{align}
By comparing the QCA indices of the two sides, we get $\delta n_2 = 0$, therefore $n_2\in Z^2(BG_b,\Z_2)$. Different circuit decomposition of $U(g)$ can shift $U_A(g)$ by multiplication of boundary 1D fermionic QCA (FQCA), which shifts $n_2$ by a coboundary $n_2\to n_2 + \delta \chi$ with $\chi\in C^1(BG_b,\Z_2)$. Therefore the anomaly index is valued in
\begin{align}
    [n_2]\in H^2(BG_b, \Z_2)~.
\end{align}
We call this anomaly the Majorana layer. We note that anomaly indices associated with nontrivial QCA indices localized at junctions of symmetry defects have also been discussed in Refs.~\cite{Tu2026anomalies,shirley2025QCA,kapustin2025highersymmetries, Feng2026onsite}.

\subsection{Complex fermion layer}
\label{subsec:complex layer}
We define subsequent anomaly indices when the Majorana layer is trivial, $[n_2]=0$. Then one can choose a truncation of $\{U_A(g)\}$ such that $n_2$ is vanishing as a cochain, so that $\{\Omega_{\partial A}(g,h)\}$ carry bosonic QCA indices: $\text{Ind}_F(\Omega_{\partial A}(g,h))\in\mathbb{Q}_+$~\cite{Tu2026anomalies, shirley2025QCA, kapustin2025highersymmetries}. This QCA index defines a cohomology class $[\nu_2]\in H^2(BG_b,\mathbb{Q}_+)$. Unlike the fermionic QCA index captured by $[n_2]$, the bosonic index $[\nu_2]$ does not lead to dynamical consequence~\cite{shirley2025QCA}, therefore we do not treat this index as an 't Hooft anomaly. However, the bosonic QCA index $[\nu_2]$ obstructs onsite realization of the symmetry and gauging on the lattice.

Let us tensor $U(g)$ with the anomaly-free $G_b$ symmetry $U'_b(g)$ in the bosonic ancillary Hilbert space $\mathcal{H}'_b$ carrying the inverse index $\nu'_2=\nu_2^{-1}$ of $U(g)$. Such anomaly-free symmetry operator in the bosonic tensor product Hilbert space has been constructed in \cite{shirley2025QCA} for generic $\nu_2$.
Then, the $G_f$ symmetry $U(g)\otimes U'_b(g)$ acting on the enlarged Hilbert space has the trivial QCA index, therefore $\{\Omega_{\partial A}(g,h)\}$ for $U(g)\otimes U'_b(g)$ is FDQC.

From now, we rewrite $U(g)\otimes U'_b(g)$ in the enlarged Hilbert space by $U(g)$ for simplicity, and study the anomaly indices of this $G_f$ symmetry. Since $\Omega_{\partial A}(g,h)$ for $U(g)$ has even fermion parity, it satisfies
\begin{align}
    \Omega_{\partial A}(g,h) \Omega_{\partial A}(gh,k) = \ ^g\left[\Omega_{\partial A}(h,k) \right]\Omega_{\partial A}(g,hk)~.
\end{align}
Since $\Omega_{\partial A}(g,h)$ is FDQC, one can truncate it to the interval $I$ of $\partial A$, which we denote by $\Omega_{I}(g,h)$. Then define the operator
\begin{align}
    \Gamma(g,h,k) = \Omega_{I}(g,h) \Omega_{I}(gh,k) \left({}^g\left[\Omega_{I}(h,k) \right]\Omega_{I}(g,hk) \right)^{-1}
\end{align}
which is supported at the ends of the interval $I$, therefore has the form of $\Gamma=\Gamma_l\Gamma_r$ where $l,r$ are the left, right end of the interval. The second anomaly index is given by fermion parity of $\Gamma_l(g,h,k)$:
\begin{align}
    n_3(g,h,k) = \begin{cases}
0, &  \text{$\Gamma_l(g,h,k)$ is fermion parity even},\\
1, & \text{$\Gamma_l(g,h,k)$ is fermion parity odd}.
\end{cases}
\end{align}

We use the notation
\begin{align}
{}^g X
&:=U_A(g)XU_A(g)^{-1},\\
{}^{(g,h)}X
&:=\Omega_I(g,h)X\Omega_I(g,h)^{-1},
\end{align}
and use concatenated superscripts for the corresponding composed
conjugations. In particular, ${}^{^g(h,k)}X$ denotes conjugation by
${}^g\Omega_I(h,k)$, and ${}^{g\cdot h}X$ denotes conjugation by
$U_A(g)U_A(h)$.

The operator $\Gamma$ obeys the non-Abelian $3$-cocycle relation
\begin{align}
&\Gamma(g,h,k)\,
{}^{^g(h,k)}\Gamma(g,hk,l)\,
{}^g\Gamma(h,k,l)
\nonumber\\
&\hspace{1cm}
=
{}^{(g,h)}\Gamma(gh,k,l)\,
\Delta_I(g,h;k,l)\,
{}^{g\cdot h(k,l)}\Gamma(g,h,kl)~,
\label{eq:nonabelian-3-cocycle}
\end{align}
where we introduced an operator
\begin{align}
\Delta_I(g,h;k,l)
:=
{}^{(g,h)\cdot gh}\Omega_I(k,l)
\left({}^{g\cdot h}\Omega_I(k,l)\right)^{-1}.
\end{align}
It is a product of local operators at the ends of $I$, which can be factorized as
\begin{align}
\Delta_I(g,h;k,l)
=
\Delta_l(g,h;k,l)\Delta_r(g,h;k,l)~.
\end{align}
Here, the operators $\Delta_l(g,h;k,l), \Delta_r(g,h;k,l)$ are fermion parity even. This can be seen by expressing $\Omega_I = \Omega_{l}\Omega_{I'}$, where $\Omega_{l}$ is a finite depth circuit acting in the vicinity of the left end $l$, and $\Omega_{I'}$ is a FDQC supported at the interval $I'\subset I$ such that the left end of $I'$ is slightly inside $I$. Using this expression, one can write $\Delta_I$ in the form of
\begin{align}
    \Delta_I(g,h;k,l) = V_l \times {}^{(g,h)\cdot gh}\Omega_{I'}(k,l)
\left({}^{g\cdot h}\Omega_{I'}(k,l)\right)^{-1},
\end{align}
where $V_l$ is fermion parity even. 
Since the end of $I'$ is inside $I$, the operator ${}^{(g,h)\cdot gh}\Omega_{I'}(k,l)
\left({}^{g\cdot h}\Omega_{I'}(k,l)\right)^{-1}$ is only supported at the right end. Therefore $\Delta_l\propto V_l$, hence fermion parity even.

Taking the fermion parity of the left-end part of
Eq.~\eqref{eq:nonabelian-3-cocycle}, we get
\begin{align}
0
&=
n_3(h,k,l)
+n_3(gh,k,l)
+n_3(g,hk,l)
\nonumber
+n_3(g,h,kl)
+n_3(g,h,k)
\nonumber\\
&=
(\delta n_3)(g,h,k,l)
\qquad \text{mod }2.
\end{align}
Therefore
\begin{align}
n_3\in Z^3(BG_b,\mathbb{Z}_2)~,
\end{align}

The cochain $n_3$ depends on the choice of interval truncation.
Suppose that the truncation is changed by endpoint-supported
operators,
\begin{align}
\widetilde{\Omega}_I(g,h)
=
\Lambda_l(g,h)\Lambda_r(g,h)\Omega_I(g,h)~,
\end{align}
where the total operator $\Lambda_l\Lambda_r$ is fermion-parity even.
Writing fermion parity of $\Lambda_l(g,h)$ by a 2-cochain $\chi(g,h)$, one finds that $n_3$ is shifted by $n_3\to n_3+\delta \chi_2$.
Consequently, after the trivialization of the Majorana layer, the anomaly index is
\begin{align}
[n_3]\in H^3(BG_b,\mathbb{Z}_2)~.
\end{align}
We call this anomaly the complex fermion layer.

\subsection{Bosonic layer}

\begin{figure}[tbh]
\centering
\includegraphics[width=0.25\textwidth]{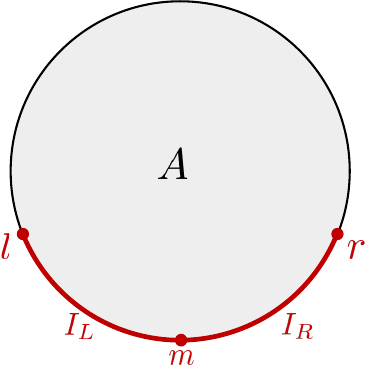}
\caption{A disk region $A$ and an interval $I$ (thick red line), separated into $I_L=[l,m]$ and $I_R=[m,r]$.
}
\label{fig:area}
\end{figure}

The left-end part of Eq.~\eqref{eq:nonabelian-3-cocycle} is only satisfied up to phase factor, which defines
$e^{2\pi i \nu_4(g,h,k,l)}\in U(1)$ by
\begin{align}
e^{2\pi i \nu_4(g,h,k,l)}
:={}& \Gamma_l(g,h,k)\,
{}^{g(h,k)}\Gamma_l(g,hk,l)\,
{}^g\Gamma_l(h,k,l)
\nonumber\\
&\times
\Bigl[
{}^{(g,h)}\Gamma_l(gh,k,l)\,
\Delta_l(g,h;k,l)\,
{}^{g\cdot h(k,l)}\Gamma_l(g,h,kl)
\Bigr]^{-1}.
\label{eq:alpha4-definition}
\end{align}
This $\nu_4$ index has been introduced for bosonic symmetries in \cite{kawagoe2025anomaly}.
The operator on the right-hand side is supported near a single
endpoint and lies in the center of the local operator algebra, and
hence is a $U(1)$ phase. In the above definition of $\nu_4$, we are fixing a canonical definition of $\Delta_l$ operator as follows:
choose a point $m$ in the interior of $I$ and write
\begin{align}
I_{L}=[l,m],
\qquad
I_{R}=[m,r].
\end{align}
See Fig.~\ref{fig:area}.
Since $\Omega_I(k,l)$ is FDQC, choose a
factorization
\begin{align}
\Omega_I(k,l)
=
\Omega_{L}(k,l)\Omega_{R}(k,l),
\label{eq:Omega-LR-factorization}
\end{align}
where $\Omega_{\mathrm L}(k,l)$ and
$\Omega_{\mathrm R}(k,l)$ are FDQCs supported at $I_{L}$ and $I_{R}$, respectively.
We define
\begin{align}
\Delta_{l}(g,h;k,l)
:=
{}^{(g,h)\cdot gh}\Omega_{L}(k,l)
\left(
{}^{g\cdot h}\Omega_{L}(k,l)
\right)^{-1}~.
\label{eq:canonical-Delta-L}
\end{align}
The two conjugation actions agree in the interior of $I$, so the
operator in Eq.~\eqref{eq:canonical-Delta-L} is supported only near
the left endpoint $l$. When the $[n_3]=0$ in cohomology, one can choose the truncation for $\Omega_I$ such that $\{\Gamma_l(g,h,k)\}$ are fermion parity even. 

Once we choose such truncations such that $\{\Gamma_l(g,h,k)\}$ are even, following the derivation of \cite{kawagoe2025anomaly}, one can then see that $\nu_4$ is closed: $\delta \nu_4 = 0$. Since the decomposition $\Gamma=\Gamma_l\Gamma_r$ to define $\Gamma_l$ has phase ambiguity $\Gamma_l(g,h,k)\to \Gamma_l(g,h,k) \cdot e^{2\pi i\phi(g,h,k)}$, this shifts $\nu_4$ by a coboundary $\nu_4\to \nu_4+\delta \phi$. Therefore the anomaly index with $n_2=0, n_3=0$ is given by 
\begin{align}
    [\nu_4]\in H^4(BG_b, \R/\Z)~.
\end{align}
We call this anomaly the bosonic layer.

\section{Dynamical consequence of anomalies in (2+1)D}
\label{sec:dynamical}
\subsection{Obstruction to SRE}

\subsubsection{Majorana layer}

Suppose that an SRE state $\ket{\Psi}=V\ket{0}$ is $G_f$ symmetric: $U(g)\ket{\Psi}\propto \ket{\Psi}$, where $V$ is FDQC and $\ket{0}$ is a product state with Fock vacuum of fermions. We show that $G_f$ symmetry then must have trivial $[n_2]\in H^2(BG_b,\Z_2)$  index, therefore the nontrivial $n_2$ index forbids SRE. 

Let us first conjugate the symmetry operator $U(g)$ by $V$, which defines another $G_f$ symmetry operator $\tilde U(g):= V^\dagger U(g) V$. Since the FQCA index of $\Omega_{\partial A}$ is invariant under the conjugation action of $V$, the new symmetry $\tilde U(g)$ carries the same $n_2$ index. The symmetry $\tilde U(g)$ preserves the product state: $\tilde U(g)\ket{0}=\ket{0}$. 

Let us consider a truncation of $\tilde U(g)$ within a disk region $\tilde U_A(g)$.
Since $\tilde U(g)\ket{0}=\ket{0}$, $\tilde U_A(g)\ket{0} = \ket{0}\otimes \ket{\Phi}_{g;\partial A}$. This 1D state $\ket{\Phi}_{g;\partial A}$ is invertible. A 1D invertible state with $\Z_2^F$ symmetry is classified by $\Z_2$, which corresponds to the trivial or Kitaev's Majorana chain~\cite{Kitaev2001unpaired}. We then define a 1-cochain $\chi\in C^1(BG_b,\Z_2)$ by
\begin{align}
    \chi(g) = \begin{cases}
0, &  \text{when $\ket{\Phi}_{g;\partial A}$ is a trivial invertible state},\\
1, & \text{when $\ket{\Phi}_{g;\partial A}$ is nontrivial, i.e., equivalent to the Kitaev's Majorana chain}.
\end{cases}
\end{align}

From \eqref{eq:projective UA}, one finds the equation
\begin{align}
    \tilde U_A(g)\tilde U_A(h)\tilde U^\dagger_A(gh)\ket{0} = \Omega_{\partial A}(g,h)\ket{0}~,
\end{align}
where we used the fact that $(-1)^{F_A}$ preserves the product state $\ket{0}$. On one hand, the state $\tilde U_A(g)\tilde U_A(h)\tilde U^\dagger_A(gh)\ket{0}$ has the form of 
\begin{align}
    \tilde U_A(g)\tilde U_A(h)\tilde U^\dagger_A(gh)\ket{0} = \ket{0}\otimes \ket{\Phi}_{(g,h);\partial A}~,
\end{align}
where $ \ket{\Phi}_{(g,h);\partial A}$ is 1D invertible state along the vicinity of $\partial A$. 
Using the cochain $\chi$, the 1D state $\ket{\Phi}_{(g,h);\partial A}$ is within the phase of $\delta\chi(g,h) = \chi(g)+\chi(h)+\chi(gh)\in \Z_2$ in the $\Z_2$ classification. On the other hand, using the expression
\begin{align}
    \Omega_{\partial A}(g,h)\ket{0} = \ket{0}\otimes \ket{\Phi}_{(g,h);\partial A}~,
\end{align}
the 1D state $\ket{\Phi}_{(g,h);\partial A}$ is within the phase of $n_2(g,h)$ in the $\Z_2$ classification, since the FQCA with the $\sqrt{2}$ index acting on a trivial invertible phase becomes the Kitaev's Majorana chain. Therefore we get
\begin{align}
    n_2=\delta \chi~,
\end{align}
i.e., the cohomology class $[n_2]\in H^2(BG_b,\Z_2)$ for the $G_f$ symmetry $U(g)$ must be trivial.

\subsubsection{Complex fermion layer}

Suppose that the Majorana layer is trivial, $n_2=0$ as a cochain, and further that the bosonic QCA indices of $\Omega_A(g,h)$ are trivialized by non-anomalous symmetry of additional ancillary Hilbert space, as discussed in Sec.~\ref{subsec:complex layer}. 
In this setup, suppose that the symmetry $U(g)$ preserves the SRE state, i.e., the conjugated symmetry $\tilde U(g)$ preserves the product state with the Fock vacuum $\ket{0}$. We will then show that $G_f$ symmetry must have trivial $[n_3]\in H^3(BG_b,\Z_2)$ index, therefore the nontrivial $n_3$ index forbids SRE. 

Because $n_2=0$, the 1D state $\ket{\Phi}_{g;\partial A}$ appearing in the symmetry action on the product state $\tilde U_A(g)\ket{0}=\ket{0}\otimes \ket{\Phi}_{g;\partial A}$ becomes a trivial invertible phase, hence SRE. Therefore, writing the disentangler of the 1D state $\ket{\Phi}_{g;\partial A}$ as $\Sigma_{\partial A}(g)$, the modified truncation of the symmetry $U'_A(g):=\Sigma_{\partial A}(g)\tilde U_A(g)$ satisfies $U'_A(g)\ket{0} = \ket{0}$.
Then, the 1D operator $\Omega'_{\partial A}(g,h)$ obtained from the modified truncation $U'_A(g)$ satisfies $\Omega'_{\partial A}(g,h)\ket{0}=\ket{0}$. Since $\Omega'_{\partial A}(g,h)$ is FDQC, its truncation to the interval $\Omega'_{I}(g,h)$ acts on the product state $\ket{0}$ in the form of
\begin{align}
    \Omega'_{I}(g,h)\ket{0} = \ket{0}\otimes \ket{\Phi}_{l;(g,h)}\otimes \ket{\Phi}_{r; (g,h)}~,
\end{align}
where $\ket{\Phi}_{l;(g,h)}, \ket{\Phi}_{r;(g,h)}$ are 0D states localized at the left, right ends of the interval $I$.
Let us write the fermion parity of $\ket{\Phi}_{l;(g,h)}$ by a 2-cochain $\chi(g,h)$.
Then, using the following definition of $\Gamma$ 
\begin{align}
    \Gamma(g,h,k) = \Omega'_{I}(g,h) \Omega'_{I}(gh,k) \left({}^g\left[\Omega'_{I}(h,k) \right]\Omega'_{I}(g,hk) \right)^{-1}~,
\end{align}
consider the action of the operators in rhs on the product state, 
\begin{align}
    \Omega'_I(g,h)\Omega'_{I}(gh,k) \left({}^g\left[\Omega'_{I}(h,k) \right]\Omega'_{I}(g,hk) \right)^{-1}\ket{0} = \ket{0}\otimes \ket{\Phi}_{l;(g,h,k)}\otimes \ket{\Phi}_{r; (g,h,k)}~.
\end{align}
Then the fermion parity of $\ket{\Phi}_{l;(g,h,k)}$ is given by $\delta\chi(g,h,k)$. Meanwhile, the action of $\Gamma$ is given by
\begin{align}
    \Gamma(g,h,k)\ket{0} = \ket{0}\otimes \ket{\Phi}_{l;(g,h,k)}\otimes \ket{\Phi}_{r; (g,h,k)}~.
\end{align}
Due to the definition of $n_3$, the fermion parity of $\ket{\Phi}_{l;(g,h,k)}$ is given by $n_3(g,h,k)$. Equating the fermion parity of $\ket{\Phi}_{l;(g,h,k)}$ gives $n_3=\delta \chi$, therefore $[n_3]\in H^3(BG_b,\Z_2)$ must be trivial.

\subsubsection{Bosonic layer}
Suppose further that $n_3=0$ as a cochain. We show that $G_f$ symmetry must have trivial $[\nu_4]\in H^4(BG_b, \R/\Z)$ index, therefore the nontrivial $\nu_4$ index forbids SRE. This statement for bosonic layer has been essentially established in \cite{kawagoe2025anomaly}, but let us present the proof for completeness. 

Because $n_3=0$, the 0D state $\ket{\Phi}_{l;(g,h)}$ appearing in the expression
\begin{align}
    \Omega'_{I}(g,h)\ket{0} = \ket{0}\otimes \ket{\Phi}_{l;(g,h)}\otimes \ket{\Phi}_{r; (g,h)}
\end{align}
becomes fermion parity even. Let us write a local unitary $\Sigma_l, \Sigma_r$ satisfying $\Sigma_l\ket{0}=\ket{\Phi}_{l;(g,h)}, \Sigma_r\ket{0}=\ket{\Phi}_{r;(g,h)}$. Then, redefining a new truncation $\Omega''_I(g,h)$ by $\Omega''_I=\Sigma_l^\dagger \Sigma_r^\dagger\Omega'_I$, $\Omega''_I$ preserves the product state: $\Omega''_I(g,h)\ket{0}=\ket{0}$. Using this $\Omega''_I$, $\Gamma$ also preserves the product state: $\Gamma(g,h,k)\ket{0}=\ket{0}$. 

In this setup, the 0D operator $\Gamma_l$ acts by a phase, $\Gamma_l(g,h,k)\ket{0} = e^{2\pi i \phi(g,h,k)}\ket{0}$. Since one can choose $\Omega_L$ in the decomposition $\Omega_I=\Omega_L\Omega_R$ such that $\Omega_L\ket{0}=\ket{0}$, using this choice of $\Omega_L$ one can see that $\Delta_l(g,h;k,l)\ket{0}=\ket{0}$. This implies that $\nu_4$ is a coboundary: $\nu_4=\delta \phi_3$, therefore $[\nu_4]\in H^4(BG_b, \R/\Z)$ must be trivial.

\subsection{Obstruction to invertible state when $\omega_2=0$ }

When the $G_f$ symmetry further has trivial extension class $\omega_2=0$, i.e., $G_f=G_b\times\Z_2^F$, then the nontrivial anomaly index ($n_2$, $n_3$ or $\nu_4$) forbids not only symmetric SRE but generic symmetric invertible state. This can be immediately seen as follows. Suppose that an invertible state $\ket{\Psi}$ is symmetric: $U(g)\ket{\Psi}\propto \ket{\Psi}$. Then there exists an inverse state $\ket{\Phi}$ such that $\ket{\Psi}\otimes \ket{\Phi}$ is SRE, then the SRE state $\ket{\Psi}\otimes \ket{\Phi}$ is symmetric under the $G_f=G_b\times\Z^F_2$ symmetry $U(g)\otimes \mathbbm{1}$, therefore $U(g)\otimes \mathbbm{1}$, hence $U(g)$, must have trivial $(n_2, n_3,\nu_4)$ index. 


We note that this argument does not apply when the extension class $\omega_2$ is nontrivial. Since $U(g) \otimes \mathbb 1$ acts only on the original system, it satisfies 
\begin{equation}
    (U(g)\otimes\mathbbm 1)(U(h)\otimes\mathbbm 1)
    = \big[(-1)^F\otimes\mathbbm 1\big]^{\omega_2(g,h)}\,
      U(gh)\otimes\mathbbm 1 ,
\end{equation}
where $(-1)^F\otimes\mathbbm 1$ is the fermion parity of the original system alone, rather than the total fermion parity $(-1)^F\otimes(-1)^{F'}$ of the enlarged system including the fermions of $|\Phi\rangle$. Hence $\{ U(g) \otimes \mathbb 1\}$ does not realize $G_f$ as a symmetry of the enlarged fermionic system in which $\mathbb Z_2^F$ must be generated by the total fermion parity. When $\omega_2 = 0$, on the other hand, this issue does not arise because $\{ U(g) \otimes \mathbb 1\}$ is a linear representation of $G_b$ that commutes with the total fermion parity $(-1)^F\otimes(-1)^{F'}$, so the two together realize $G_b\times\mathbb Z_2^F$ on the enlarged system, with $\mathbb Z_2^F$ generated by the total fermion parity.

Nevertheless, for the $\mathbb Z_4^F$ symmetry $U(g)$ discussed shortly in Sec.~\ref{sec:anomaly free sym forbids SRE}, a restricted version of this argument applies. Instead of $U(g) \otimes \mathbb 1$, we act on the englarged system with $U(g) \otimes V(g)$, where $V(g)$ is an ordinary onsite $\mathbb Z_4^F \subset U(1)_f$ symmetry of the ancilla such that $(U(g) \otimes V(g))^2 = (-1)^F$ with $(-1)^F$ the total fermion parity. This requires the inverse state $\ket \Phi$ to be symmetric under an onsite $\Z_4^F$ symmetry, e.g., a stack of Chern insulators, which is possible only when $\ket \Psi$ has $c_- \in \mathbb Z$. The nontrivial index $n_2 = \omega_2$ therefore forbids symmetric invertible states with $c_-\in\mathbb Z$.

\section{Dynamical consequence of a symmetry with trivial QFT anomaly}
\label{sec:anomaly free sym forbids SRE}

\subsection{$\Z_4^F$ lattice symmetry and dynamical consequence}

\paragraph{Obstruction to SRE.}
Here we consider $\Z_2^F\to \Z_4^F\to \Z_2$ symmetry, generated by an operator $U(g)$ satisfying $U(g)^2=(-1)^F$.
The extension class $\omega_2\in H^2(B\Z_2,\Z_2)=\Z_2$ is nontrivial.
We will explicitly construct an example of the lattice $\Z_4^F$ symmetry with nontrivial $[n_2]\in H^2(B\Z_2,\Z_2)$ index.
This is obtained by a slight modification of the symmetry operator constructed in \cite{shirley2025QCA} as follows.

\begin{figure}[tbh]
\centering
\includegraphics[width=0.65\textwidth]{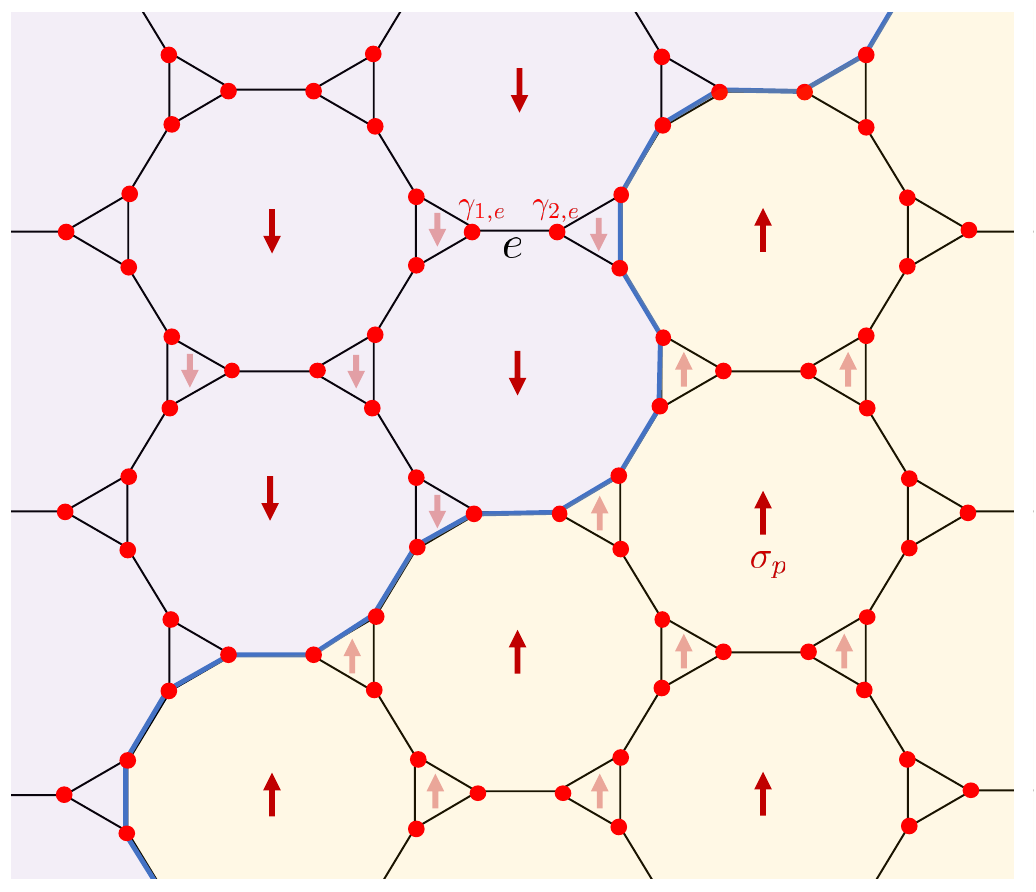}
\caption{The lattice $\Gamma$. Its edges will be endowed with Kasteleyn orientation. Each vertex of $\Gamma$ has a Majorana fermion (red dot), and each honeycomb lattice plaquette $p$ has a qubit $\sigma_p$. We also place a fictitious qubit on each triangle plaquette, whose state is fixed by the majority rule of three qubits surrounding the triangle. The $\Z_2$ domain wall forms a closed loop (thick blue line).
}
\label{fig:Gammalattice}
\end{figure}

We consider a honeycomb lattice with a qubit $\sigma_p$ on each plaquette $p$ and two Majorana fermions $\gamma_{e,1}$ and $\gamma_{e,2}$ on each edge $e$. 
The Majorana sites, together with the links connecting neighboring Majorana fermions at a common honeycomb vertex, form a graph $\Gamma$: see Fig.~\ref{fig:Gammalattice}.
We also introduce a fictitious qubit $\sigma_p$ at each triangle face of $\Gamma$; the state of the fictitious qubit is fixed by $\sigma_p=+1$ or $-1$ depending on whether the majority of three qubits surrounding the triangle has $+1$ or $-1$.

We fix a Kasteleyn orientation on the edges of $\Gamma$, i.e., an orientation such that every even-length cycle in $\Gamma$ contains an odd number of edges oriented counterclockwise. For every oriented link between two neighboring Majorana sites $i$ and $j$, we define
\begin{equation}
    s_{ij}=
    \begin{cases}
        +1, & \text{if the Kasteleyn arrow points from $i$ to $j$},\\
        -1, & \text{if the Kasteleyn arrow points from $j$ to $i$},
    \end{cases}
    \qquad s_{ij}=-s_{ji}.
\end{equation}

For each honeycomb edge $e$, let $s_e\equiv s_{(e,1)(e,2)}$ denote the Kasteleyn sign on the link connecting its two Majorana fermions. We combine these Majorana fermions into a local complex fermion,
\begin{equation}
    c_e=\frac{1}{2}\left(\gamma_{e,1}-i s_e\gamma_{e,2}\right),
    \qquad
    n_e=c_e^\dagger c_e.
\end{equation}
With this convention, the fermion parity associated with edge $e$ is
\begin{equation}
    P_e=(-1)^{n_e}=i s_e\gamma_{e,1}\gamma_{e,2},
\end{equation}
and the total fermion parity is
\begin{equation}
    (-1)^F=\prod_e P_e=\prod_e e^{i\pi n_e}.
\end{equation}

Let $p(e)$ and $q(e)$ be the two plaquettes adjacent to edge $e$, and define the domain-wall projector
\begin{equation}
    w_e=\frac{1-\sigma_{p(e)}^z\sigma_{q(e)}^z}{2}.
\end{equation}
Thus, $w_e=1$ when $e$ lies on a domain wall of the plaquette spins and $w_e=0$ otherwise. For every fixed configuration of the $\sigma_p^z$ eigenvalues, the edges with $w_e=1$ form closed loops for the $\Z_2$ domain wall.

We orient every domain-wall loop so that the $\sigma^z=+1$ domain lies to its left and the $\sigma^z=-1$ domain lies to its right. Let $\widehat{\mathcal T}[\{\sigma_p^z\}]$ denote the QCA that translates the Majorana fermions by one Majorana site in the forward direction along every oriented domain-wall loop. More explicitly, if $i+1$ is the Majorana site following $i$ along an oriented domain wall, then
\begin{equation}
    \widehat{\mathcal T}^{-1}\gamma_i\widehat{\mathcal T}
    =s_{i,i+1}\gamma_{i+1}.
\end{equation}
The unitary $\widehat{\mathcal T}$ acts trivially on Majoranas that do not lie on a domain wall.

The generator of the fermionic symmetry is defined directly by
\begin{equation}
    U_g
    =
    \widehat{\mathcal T}[\{\sigma_p^z\}]
    \prod_e
    \exp\!\left[\frac{\pi i}{2}(1-w_e)n_e\right]
    \prod_p\sigma_p^x
    \label{eq:z4f-generator}
\end{equation}
The three factors in Eq.~\eqref{eq:z4f-generator} have the following interpretation. The last factor flips all plaquette spins. The first factor translates the Majoranas along the resulting domain walls. The middle factor acts only on edges that are not on a domain wall and implements a quarter rotation of the corresponding complex fermion. 

Let us verify the $\mathbb{Z}_4^F$ algebra of the symmetry. The global spin flip leaves the unoriented domain-wall configuration invariant,
\begin{equation}
    w_e[\{-\sigma_p^z\}]=w_e[\{\sigma_p^z\}],
\end{equation}
but reverses the orientation of every domain-wall loop. Therefore, in the square of $U_g$, the two translation operators move the Majorana operators in opposite directions. A Majorana $\gamma_i$ on a domain wall is translated to its neighbor and then back to its original position, acquiring the Kasteleyn sign
\begin{equation}
    \gamma_i
    \longmapsto
    s_{i,i+1}s_{i+1,i}\gamma_i
    =-\gamma_i.
\end{equation}
That is, $U_g^2$ acts on the fermions on the domain wall by fermion parity, as demonstrated in \cite{shirley2025QCA}.

On an edge away from the domain walls, each action of $U_g$ instead performs a $\pi/2$ number rotation on fermions. Square action therefore gives
\begin{equation}
    \left(e^{\pi i n_e/2}\right)^2
    =e^{i\pi n_e}
    =(-1)^{n_e}.
\end{equation}
Hence $U_g^2$ acts by fermion parity away from the domain walls as well. Therefore we get the $\Z_4^F$ symmetry algebra
\begin{align}
    U_g^2= (-1)^F~.
\end{align}

Due to the Majorana translation on the $\Z_2$ domain wall, the truncated symmetry operator within a disk region $U_A(g)$ has the algebra
\begin{align}
    (U_A(g))^2 = \Omega_{\partial A} \times (-1)^{F_A}~,
\end{align}
where $\Omega_{\partial A}$ is the Majorana translation with the FQCA index $\sqrt{2}$. Therefore the $\Z_4^F$ symmetry carries nontrivial $n_2$ index,  given by $n_2=\omega_2$ using the extension class $[\omega_2]\in H^2(B\Z_2,\Z_2)$.
Thus, the $\Z_4^F$ symmetry forbids an SRE state. 

\paragraph{Obstruction to invertible state with $c_-\in\Z$.}

Here we further show that the $\mathbb Z_4^F$ symmetry $U(g)$ obstructs generic invertible states with integer chiral central charge $c_-\in\mathbb Z$.
The continuum classification of (2+1)D invertible phases with \textit{onsite} $\mathbb Z_4^F$ symmetry is $\mathbb Z$, characterized by the integer chiral central charge $c_-\in\mathbb Z$~\cite{barkeshli2021invertible, kobayashi2026remarks}. Therefore, our result implies that $U(g)$ does not preserve any invertible state compatible with the onsite $\mathbb Z_4^F$ symmetry. 

Suppose that the lattice $\mathbb{Z}_4^F$ symmetry $U(g)$ 
admitted a symmetric invertible state $\ket{\Phi}$ with $c_-=k\in\mathbb Z$. 
Stacking it with an ordinary onsite $\mathbb{Z}_4^F$ symmetric invertible state $\ket{\Psi_{-k}}$ of chiral central charge $c_-=-k$ (e.g., $k$ copies of the free fermion Chern insulator with charge $\Z_4^F\subset U(1)^f$ symmetry) produces a symmetric invertible state with vanishing total chiral central charge $\ket{\Phi}\otimes \ket{\Psi_{-k}}$, which is expected to be SRE.\footnote{Such state is generally not a strict SRE, but disentangled by FDQC with exponentially decaying tails of each local circuit. Since the classification of 1D (F)QCA is robust against these tails \cite{ranard2026approximateqcasdimensionusing, Ranard2022converse}, we expect that the anomaly indices forbid such SRE states with tails as well.
We emphasize that the absence of strict locality introduces additional subtleties in the precise definition of these anomaly indices, and we do not attempt a fully rigorous treatment in the present work.
}

Since the FQCA index is multiplicative under tensor product (and therefore the Majorana-layer index is additive under stacking) the onsite symmetry does not modify the lattice anomaly index $n_2=\omega_2$, and the symmetry of the stacked state still carries the same nontrivial $n_2$. This obstructs symmetric SRE, and contradicts the realization with the SRE state $\ket{\Phi}\otimes \ket{\Psi_{-k}}$. Therefore, the lattice symmetry with $n_2=\omega_2$ is incompatible not only with $c_-=0$ SRE states, but with any symmetric invertible state with $c_-\in\Z$. 

This conclusion should be contrasted with an onsite realization of the same group. Every invertible state symmetric under an onsite $\mathbb Z_4^F$ symmetry has $c_-\in\mathbb Z$, and every integer value is attained by stacks of Chern insulators~\cite{barkeshli2021invertible, kobayashi2026remarks}.
Thus, $U(g)$ is incompatible with every invertible phase that an onsite $\mathbb Z_4^F$ symmetry $V(g)$ supports. This distinction is visible only in the lattice anomaly index, $n_2=\omega_2$ for $U(g)$ versus $n_2=0$ for $V(g)$: the continuum 't~Hooft anomaly of $\mathbb Z_4^F$ is trivial, so both symmetry actions descend to the same, trivial continuum anomaly. Non-onsiteness alone does not produce this effect: a symmetry of the form $W^\dagger V W$ with $V$ onsite and $W$ an FDQC, or one carrying only a bosonic QCA index, supports the same invertible phases as an onsite symmetry. What distinguishes $U(g)$ is its Majorana-layer index $n_2=\omega_2$. This index does not exclude invertible phases altogether: a lattice $\mathbb Z_4^F$ symmetry with the same index is realized exactly in a $p+ip$ superconductor by a quasi-local operator~\cite{kobayashi2026remarks}, and stacking with onsite-symmetric Chern insulators yields symmetric invertible states with every $c_-\in\mathbb Z+1/2$.

\subsection{$\Z_4^F$ symmetry in (2+1)D has trivial QFT anomaly}
In (2+1)D continuum QFT, the 't Hooft anomaly of the $\Z_4^F$ symmetry is trivial~\cite{WangGu2020,Garcia_Etxebarria_2019}. 
The continuum QFT anomaly is labeled by a triple $(n_2, n_3, \nu_4)\in Z^2(BG_b,\Z_2)\times C^3(BG_b, \Z_2)\times C^4(BG_b, \R/\Z)$ satisfying a set of consistency equations among the cochains~\cite{WangGu2020}. For $G_f=\mathbb{Z}_4^F$, a representative with $n_2=\omega_2$ is allowed by these consistency conditions. However, the anomaly classification is defined modulo equivalences of the triple, including
$n_2\sim n_2 + \omega_2$ associated with equivalences of $n_3, \nu_4$ accordingly.\footnote{In the (3+1)D bulk fermionic SPT describing the inflow, the equivalence $n_2\sim n_2+\omega_2$ can be understood as a symmetry action in the dual (3+1)D $\mathbb{Z}_2$ gauge theory obtained by bosonizing the fermionic SPT. The corresponding automorphism is implemented by pumping a $p+ip$ state and induces a permutation action on the 1-form symmetries~\cite{Barkeshli:2023bta}. In particular, it acts on a magnetic surface operator by dressing the condensation defect of a fermionic particle. This operation relabels the 2-form background fields for the 1-form symmetries, precisely corresponding to the transformation $n_2\to n_2+\omega_2$~\cite{Barkeshli:2022edm}. In the anomalous (2+1)D theory, this symmetry action corresponds to adding a counterterm associated with a $p+ip$ state, which preserves the $\mathbb{Z}_4^F$ symmetry with the lattice anomaly $n_2=\omega_2$ \cite{kobayashi2026remarks}.}
Consequently, an anomaly representative with $n_2=\omega_2$ is identified with one having a trivial $n_2$ component in the continuum QFT classification.

On the lattice, the $\mathbb{Z}_4^F$ symmetry considered above carries $n_2=\omega_2$ and forbids symmetric invertible states with $c_-\in\Z$. However, this does not imply that the $\Z_4^F$ symmetry forbids any invertible state altogether: the symmetry still preserves an invertible state with $c_-\in \Z+1/2$ \cite{kobayashi2026remarks}. 
Thus, the lattice anomaly index $n_2=\omega_2$ leads to nontrivial dynamical consequence of shifting the chiral central charge $c_-$ of the invertible state by 1/2, even though such dynamical consequence is discarded under the equivalence relation used in the continuum anomaly classification.

\section{Discussions and conjectures}

\label{sec:discussion}

In this work, we developed a lattice characterization of finite internal fermionic symmetries in (1+1)D and (2+1)D in terms of obstructions to symmetric invertible states. A general observation is that the dynamical consequence extracted from an exact lattice symmetry need not coincide with the full anomaly classification of continuum fermionic QFT.

In (1+1)D, exact lattice symmetries realize only the $(n_2,\nu_3)$ layers and miss the additional $n_1$ layer present in the continuum. In (2+1)D, the lattice anomaly indices can instead retain information that becomes trivial under the equivalence relations of the continuum classification, while still constraining which invertible phases can preserve the symmetry. The $\mathbb{Z}_4^F$ symmetry with Majorana-layer index $n_2=\omega_2$ illustrates this: it is incompatible with symmetric invertible phases having $c_-\in\mathbb{Z}$, but compatible with those having $c_-\in\mathbb{Z}+1/2$ \cite{kobayashi2026remarks}. The index therefore acts not as an obstruction to symmetric invertible phases but as a selection rule on which ones can support the given microscopic symmetry action.

Our results motivate a classification of lattice anomalies based on obstructions to symmetric SRE states. We expect such a classification to form a group under stacking, in which two symmetry actions are identified when their difference under stacking admits a symmetric SRE state, and to admit a homomorphism
\begin{align}
\varphi_{G_f}\colon
\mathcal A^{\mathrm{SRE}}_{\mathrm{lat}}(G_f)
\longrightarrow
\mathcal A_{\mathrm{QFT}}(G_f)~.
\end{align}
Here, $\mathcal A^{\mathrm{SRE}}_{\mathrm{lat}}(G_f)$ denotes this classification for exact internal $G_f$ symmetry actions on tensor-product Hilbert spaces, with the trivial class corresponding to actions admitting symmetric SRE states, and $\mathcal A_{\mathrm{QFT}}(G_f)$ the group of continuum 't~Hooft anomalies. Ref.~\cite{Tu2026anomalies} discusses a related map using a different notion of lattice anomaly, which can remain nontrivial even when the symmetry admits a symmetric product state.

The physical interpretation of $\varphi_{G_f}$ is to associate a lattice anomaly class with the continuum anomaly inherited by the infrared theory. To motivate this assignment, consider a lattice symmetry action representing a class $a$ and a symmetric Hamiltonian admitting a continuum infrared description. When the inherited symmetry action is fully specified by a homomorphism $\Phi\colon G_f\to G_{\mathrm{IR}}$, the candidate image is $\varphi_{G_f}(a)=\Phi^*(a')$, where $a'\in\mathcal A_{\mathrm{QFT}}(G_{\mathrm{IR}})$ is the IR anomaly. The pullback expresses this anomaly in terms of the original microscopic symmetry $G_f$. More generally, symmetry fractionalization or transmutation can require describing the induced map of symmetry backgrounds, rather than only a homomorphism of symmetry groups \cite{Barkeshli2019Symmetry,Benini:2018reh,Seiberg2025transmutation}; we do not attempt to define $\varphi_{G_f}$ in that generality here.

A general construction of this framework requires establishing that the resulting continuum anomaly depends only on $a$, independently of the representative symmetry action and of the symmetric Hamiltonian chosen to realize it. Given this, one would further need to check that the assignment respects stacking, which is natural to expect since stacking two lattice systems produces an infrared theory whose anomaly is the sum. We leave these questions to future work.

Our results show that the map $\varphi_{G_f}$ is neither surjective nor injective in general. For example,
\begin{itemize}
\item For $G_f=\mathbb Z_2\times \mathbb Z_2^F$ in (1+1)D, $\varphi_{G_f}$ is not surjective: QFT anomalies with odd classes modulo $8$ cannot be realized by an exact $\mathbb Z_2$ symmetry on the lattice.
\item For $G_f=\mathbb Z_4^F$ in (2+1)D, $\varphi_{G_f}$ is not injective: a nontrivial lattice obstruction to symmetric SRE states is mapped by $\varphi_{G_f}$ to a trivial continuum QFT anomaly.
\end{itemize}

The image of $\varphi_{G_f}$ consists of continuum anomalies that admit realizations by exact microscopic symmetries on tensor-product Hilbert spaces.
Its kernel captures lattice obstructions to symmetric SRE states that become trivial in the continuum QFT description, while its cokernel captures continuum anomalies that cannot arise from an exact microscopic symmetry. It would be interesting to determine systematically the structure of $\varphi_{G_f}$ for general fermionic symmetry groups $G_f$ and in arbitrary spacetime dimensions.

In particular, an important open problem is to determine whether the anomaly indices introduced in this work provide a complete characterization of the anomalies associated with exact lattice symmetries in (2+1)D. Resolving this question would require establishing that symmetries with trivial anomaly indices is onsiteable in (2+1)D. Such a result would clarify whether there exist obstructions to symmetric invertible states beyond the $n_2,n_3,\nu_4$ indices introduced here.
In (1+1)D, for a trivial extension $G_f=G_b\times \mathbb Z_2^F$, we showed that the vanishing of $(n_2,\nu_3)$ is sufficient to render the symmetry onsite after tensoring ancillary degrees of freedom and conjugating by a finite-depth circuit. It is natural to conjecture that an analogous result holds in one higher dimension.

We therefore propose the following conjecture: 
\begin{quote}
\textbf{Conjecture 1.}
For a finite unitary internal fermionic symmetry $G_f$ in (2+1)D, the symmetry is onsiteable if and only if all lattice anomaly indices $n_2,n_3,\nu_4$, together with the bosonic QCA index $[\nu_2]\in H^2(BG_b, \mathbb{Q}_+)$,
are trivial.
\end{quote}

Establishing this conjecture would show that $n_2,n_3,\nu_4$ provide faithful characterization of the 't Hooft anomaly of an exact unitary lattice symmetry in (2+1)D. 

Regarding the above conjecture, a particularly interesting extension is to anti-unitary symmetries. For fermionic anti-unitary symmetry, the continuum anomaly contains an additional layer absent from the hierarchy discussed above. Schematically, the anomaly data take the form
$(n_1,n_2,n_3,\nu_4)$ with
$[n_1]\in H^1(BG_b,\mathbb Z_T), $ 
where $\Z_T$ denotes the $G_b$-module $\Z$ with the anti-unitary elements acting by sign reversal, $n\to -n$. The $n_1$ layer may be viewed as a $p+ip$ layer: a symmetry defect of the bulk (3+1)D SPT can carry an invertible chiral fermionic state with $c_-=1/2$~\cite{WangGu2020}. 

Suppose that an analogue of Conjecture 1 continues to hold for anti-unitary symmetries, namely that an exact lattice symmetry with trivial lattice indices $(n_2,n_3,\nu_4)$ is onsiteable. Since an onsite symmetry cannot carry a nontrivial 't Hooft anomaly, this would immediately imply that a continuum anomaly whose only nontrivial component is the $p+ip$ layer cannot arise from an exact symmetry on a tensor product Hilbert space. This motivates a second conjecture: 
\begin{quote}
\textbf{Conjecture 2.}
A fermionic 't Hooft anomaly of anti-unitary symmetry with nontrivial
$[n_1]\in H^1(BG_b,\mathbb Z_T)$
cannot be realized by an exact internal symmetry on a tensor product Hilbert space in 2D space.
\end{quote}

This conjecture is the (2+1)D analogue of the phenomenon observed in (1+1)D, where continuum anomalies with nontrivial $[n_1]\in H^1(BG_b,\mathbb Z_2)$ cannot be realized by exact lattice symmetries.
The most prominent example in (2+1)D is time-reversal symmetry satisfying
$T^2=(-1)^F,$
corresponding to the fermionic anti-unitary symmetry commonly associated with the $\mathrm{Pin}^+$ structure. In (2+1)D continuum QFT, its 't Hooft anomaly is classified by $\mathbb Z_{16}$~\cite{Kapustin2015spinbordism, Witten:2016cio, Tata2022anomalies}. The odd elements of this classification carry a nontrivial $p+ip$ layer. Conjecture 2 therefore leads to the statement
\begin{quote}
\textbf{Conjecture 3.}
The odd elements of the $\mathbb Z_{16}$ time-reversal anomaly for
$T^2=(-1)^F$
cannot be realized by an exact time-reversal symmetry on a tensor product Hilbert space in 2D space.
\end{quote}

Equivalently, if exact lattice realizations exist only when the $p+ip$ layer is trivial, the anomalies accessible to exact time-reversal symmetry on the lattice would form the even subgroup of the continuum $\mathbb Z_{16}$ classification. Realizations of the odd classes would then necessarily require a more general action of symmetry than an exact internal lattice symmetry, analogous to the emanant symmetry that realizes the missing odd anomaly classes in the (1+1)D mod 8 anomaly of $\Z_2$ symmetry.
It would be valuable to establish these conjectures directly by extending the onsiteability construction to (2+1)D.

\section*{Acknowledgments}
We thank Po-Shen Hsin for discussions.
A.C. is
supported by the National Science Foundation Graduate Research Fellowship under Grant No. DGE-2036197.
R.K. is supported by the Department of Applied Physics,
the University of Tokyo.

\bibliographystyle{utphys}
\bibliography{bibliography}

\end{document}